\documentclass{article}

\PassOptionsToPackage{sort&compress,numbers}{natbib}

\usepackage[preprint]{neurips_2025}

\usepackage[utf8]{inputenc} 
\usepackage[T1]{fontenc}    
\usepackage{hyperref}       
\usepackage{url}            
\usepackage{booktabs}       
\usepackage{amsfonts}       
\usepackage{nicefrac}       
\usepackage{microtype}      
\usepackage{xcolor}         
\usepackage{graphicx}
\usepackage{amsmath} 
\usepackage{multicol} 
\usepackage{algorithm} 
\usepackage{algpseudocode} 
\usepackage{amssymb} 
\usepackage[capitalise]{cleveref} 
\usepackage{makecell} 
\usepackage{chngcntr} 
\DeclareMathOperator*{\argmin}{argmin} 
\newcommand{\eqrefplain}[1]{Eq.~\ref{#1}}

\title{OMiSO: Adaptive optimization of state-dependent brain stimulation to shape neural population states}

\author{%
  Yuki Minai$^{1,2,3}$, Joana Soldado-Magraner$^{1,3,4}$, Byron M. Yu$^{1,3,4,5*}$, Matthew A. Smith$^{1,3,5*}$\\
  $^1$Neuroscience Institute, Carnegie Mellon University\\
  $^2$Machine Learning Department, Carnegie Mellon University\\
  $^3$Center for the Neural Basis of Cognition\\
  $^4$Department of Electrical and Computer Engineering, Carnegie Mellon University \\
  $^5$Department of Biomedical Engineering, Carnegie Mellon University \\
  \texttt{\{yminai,jsoldado,byronyu,msmith\}@andrew.cmu.edu}\\
  *Denotes equal contribution.
}

\begin{document}
\maketitle
\begin{abstract}
The coordinated activity of neural populations underlies myriad brain functions. Manipulating this activity using brain stimulation techniques has great potential for scientific and clinical applications, as it provides a tool to causally influence brain function. The state of the brain affects how neural populations respond to incoming sensory stimuli. Thus, taking into account pre-stimulation neural population activity may be crucial to achieve a desired causal manipulation using stimulation. In this work, we propose Online MicroStimulation Optimization (OMiSO), a brain stimulation framework that leverages brain state information to find stimulation parameters that can drive neural population activity toward specified states. OMiSO includes two key advances: i) it leverages the pre-stimulation brain state to choose optimal stimulation parameters, and ii) it adaptively refines the choice of those parameters by considering newly-observed stimulation responses. We tested OMiSO by applying intracortical electrical microstimulation in a monkey and found that it outperformed competing methods that do not incorporate these advances. Taken together, OMiSO provides greater accuracy in achieving specified activity states, thereby advancing neuromodulation technologies for understanding the brain and for treating brain disorders.
\end{abstract}

\section{Introduction}
Most brain functions are realized through the coordinated activity of neural populations \citep{averbeck2006neural, saxena2019towards, vyas2020computation, ebitz2021population}. Causal perturbations of the brain, for example with electrical stimulation, can influence brain function by modulating neural population activity \citep{cohen2004electrical}. This activity is variable from moment to moment, and such variability in part reflects the current state of the brain \citep{averbeck2006neural, cohen2011measuring}. The brain's state is an important factor affecting how neural populations respond to incoming sensory stimuli 
\citep{reynolds2004attentional, mccormick2020neuromodulation}, and therefore also important in understanding how the brain responds to causal perturbations.

Closed-loop brain stimulation tools enable us to find stimulation parameters to achieve targeted modulation of brain activity and/or behavior. Previous studies proposed closed-loop brain stimulation methods to find optimal brain stimulation parameters to drive neural population activity toward specified states \citep{tafazoli2020learning, minai2024miso}. These methods learn and update the mapping between a wide range of potential brain stimulation parameter combinations and brain responses during an experiment. However, they do not consider the state of the brain when choosing the stimulation parameters (but see \citep{yang2021modelling} for a study testing this idea in simulation). Previous studies in deep brain stimulation (DBS) have leveraged brain state information to update stimulation parameters for improving treatment efficacy \citep{guez2008adaptive, little2013adaptive} or reducing the energy consumption of the stimulation device \citep{gao2023offline}. These methods were designed for manipulating low-dimensional bio-markers (e.g., local field potentials \citep{gao2023offline}), rather than high-dimensional neural population activity.

\begin{figure}[!tb]
\centering
\includegraphics[scale=1]{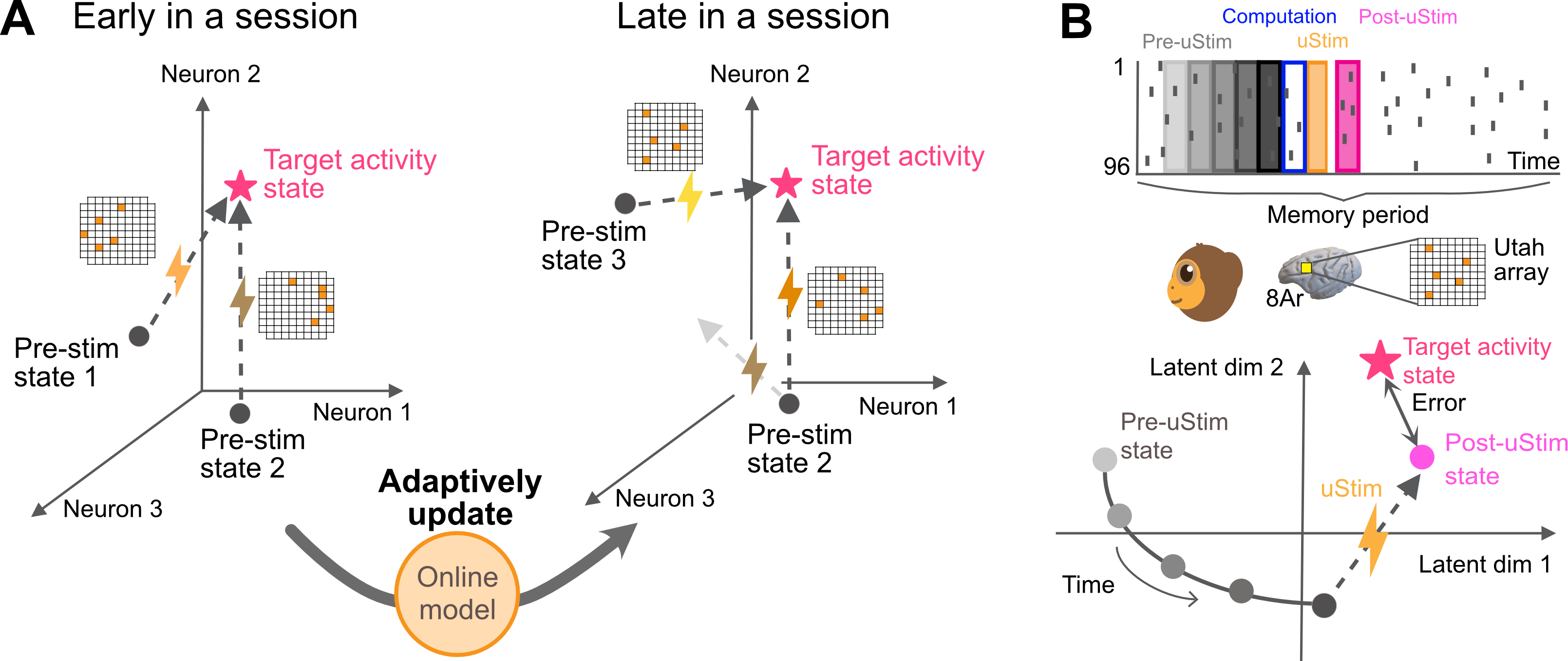}
\caption{\textbf{Goal of OMiSO and experimental paradigm.} (A) OMiSO finds brain stimulation parameters (electrode locations, orange cells) to drive neural population activity toward a target state (pink star) given a pre-stimulation state (black dots). OMiSO adaptively updates a statistical model to account for changes in pre-stimulation states (e.g., a new Pre-stim state 3) and changes in the stimulation-response mapping for a given pre-stimulation state (e.g., for Pre-stim state 2, the same stimulation parameter early in the session does not produce the target state late in the session). (B) Experimental setup. Top: The monkey performed a memory guided saccade task (Section~\ref{experimentprotocol_sup_method}) while spiking activity was recorded from a multi-electrode array implanted in PFC. Pre-uStim activities (gray bars) were recorded over five 50 ms bins. During the 50 ms computation period (blue unfilled bar), the system chose a uStim pattern, which was then applied for 40 ms (orange bar). Post-uStim activity was measured 10 ms after stimulation offset (pink bar). Bottom: neural population activity was analyzed in a low-d latent space (e.g., 2-d) of the high-d activity.}
\label{overview_fig}
\end{figure}

In this study, we develop Online MicroStimulation Optimization (OMiSO), a brain stimulation framework that uses a statistical model updated during an experiment (i.e., online) with pre-stimulation brain state information to find the stimulation parameters expected to induce a specified neural population activity. Specifically, OMiSO first fits stimulation-response models, which predict brain responses to different stimulation parameter configurations given brain states prior to stimulation (pre-uStim). OMiSO then inverts the trained stimulation-response models to output an optimal stimulation parameter configuration for creating a targeted brain state. OMiSO adaptively updates the inverse model during a stimulation experiment to improve the stimulation parameter configuration that is produced. 

We tested OMiSO using electrical microstimulation (uStim) in a monkey implanted with a multi-electrode array in the PFC (area 8Ar). OMiSO optimized the location of five stimulated electrodes applied on each trial. Taking into account pre-uStim state information was beneficial to improve the prediction of the brain's response to uStim (Section~\ref{single_trial_var_result}). An inverse model successfully predicted the stimulation parameters (electrode locations) necessary to achieve specified neural population activity states (Section~\ref{behavior_cloning_result}). By adaptively updating the inverse model using newly observed samples, OMiSO significantly improved the uStim parameter optimization performance (Section~\ref{ppo_result}).

\section{Methods}
\subsection{OMiSO overview} \label{generalframework}
The goal of OMiSO is to find a stimulation parameter configuration to drive neural population activity toward a targeted state based on pre-stimulation neural population activity states (\cref{overview_fig}A, left). OMiSO adaptively updates a statistical model to account for different pre-stimulation neural population activity states as well as changes in the stimulation-response mapping for a given pre-stimulation state (\cref{overview_fig}A, right). Specifically, OMiSO first collects and merges neural activity across multiple experimental sessions by using latent space alignment (Section~\ref{alignment_method}). The merged datasets are used to fit stimulation-response models, which predict the post-stimulation latent state given stimulation parameters and the pre-stimulation latent state (Section~\ref{model_fitting_method}). OMiSO inverts the trained stimulation-response models to output stimulation parameters that can create a target state (Section~\ref{bc_method}). The inverse model is then used to choose stimulation parameters in a brain stimulation experiment. On each trial, OMiSO analyzes a pre-stimulation latent state and chooses stimulation parameters by passing the analyzed pre-stimulation latent state and user-defined target state to the inverse model (Section~\ref{pattern_selection_method}). Using new observations, OMiSO adaptively updates the inverse model to improve the optimization performance during a brain stimulation experiment (Section~\ref{ppo_method}). In the following sections, we refer to a specific stimulation parameter configuration as a “stimulation pattern”. All hyperparameters used in the model fitting and brain stimulation experiments are summarized in Section~\ref{parameters_sup_method}.

\subsection{Latent space identification and alignment} \label{alignment_method}
When the stimulation parameter space is large, one can typically test only a small fraction of all possible stimulation patterns within an experimental session. This requires merging neural activity across sessions to create a large enough set of stimulation-response samples to learn their relationship. To merge neural activity across multiple experimental sessions, OMiSO identifies a low-dimensional (low-d) latent space of the high-dimensional (high-d) population activity in each session and aligns identified latent spaces from each experimental session. Each session consists of two types of trials: ``stimulation trials'', in which we applied stimulation, and ``no-stimulation trials'', in which we did not apply stimulation. Following \citep{minai2024miso}, we used Factor Analysis (FA) to identify the intrinsic latent space only using no-stimulation trials. For alignment, we solve the Procrustes problem to find an orthogonal transformation matrix to maximize the alignment between two FA loading matrices \citep{degenhart2020stabilization}. 

After collecting $R$ experimental sessions, for each $i$th session, OMiSO extracts a list of ${n_i}$ ``usable'' electrodes $\boldsymbol{e_i} \in \mathbb{R}^{n_i}$, where each element of $\boldsymbol{e_i}$ is an integer representing one of the electrodes.  An electrode is deemed ``usable'' if it satisfies criteria involving its mean firing rate, Fano factor, and coincident spiking with other electrodes (see Section~\ref{preprocessing_sup_method}).
For each time bin indexed by $j=1,...,J_i$ (where $J_i$ is the total number of time bins used to fit FA for the $i$th session), OMiSO takes spike counts on each usable electrode $\boldsymbol{x}_{i,j} \in \mathbb{R}^{n_i}$ and fits the following FA model using the EM algorithm:
\begin{equation}
\begin{matrix}
\boldsymbol{z}_{i,j} \sim\ \mathcal{N}(\bold{0},I) \\
\boldsymbol{x}_{i,j}|\boldsymbol{z}_{i,j} \sim\ \mathcal{N}(\Lambda_i \boldsymbol{z}_{i,j} + \boldsymbol{\mu_i},\Psi_i)  \label{eq:1}
\end{matrix}
\end{equation}
where $\boldsymbol{z}_{i,j} \in \mathbb{R}^{m}$ ($m < n_i$) is the low-d latent activity for the $j$th time bin, $\Lambda_i \in \mathbb{R}^{n_i \times m}$ is the loading matrix whose columns define the low-d latent space, $\boldsymbol{\mu_i} \in \mathbb{R}^{n_i}$ contains the mean spike counts for each electrode, and $\Psi_i \in \mathbb{R}^{n_i \times n_i}$ is a diagonal matrix capturing the independent variance of the spike counts for each electrode. 

OMiSO then defines the latent space identified in one of the $R$ sessions as a reference latent space $\Lambda_0 \in \mathbb{R}^{n_0 \times m}$ and aligns the latent space of the $i$th session $\Lambda_{i}$ to $\Lambda_{0}$. Concretely, for the $i$th session, OMiSO identifies an orthogonal transformation matrix $\hat{O_i} \in \mathbb{R}^{m \times m}$ that fulfills: 
\begin{equation}
\hat{O_i} = \argmin\limits_{O: OO^\intercal=I} \| \Lambda_{0}(\boldsymbol{e_\text{stable}}, :)-\Lambda_i(\boldsymbol{e_\text{stable}}, :) O^\intercal \|^2_F
\end{equation}
where $\boldsymbol{e_\text{stable}}$ is a list of $n_\text{stable}$ stable electrodes that are common to
the reference session and the $i$th session, identified using the method in \citep{degenhart2020stabilization}, and $\|\cdot\|_F$ is the Frobenius Norm. This optimization can be solved in closed-form \citep{schonemann1966generalized}. The $\hat{O_i}$ found is applied to $\Lambda_{i}$ to obtain the aligned latent space $\tilde{\Lambda}_{i} \in \mathbb{R}^{n_i \times m}$:
\begin{equation}
\tilde{\Lambda}_{i} = \Lambda_{i}\hat{O_i}^\intercal
\end{equation}
The latent activities in the pre-stimulation (\cref{overview_fig}B, bottom panel, gray dots) and post-stimulation periods (\cref{overview_fig}B, bottom panel, pink dot) are estimated as the posterior mean from the FA model (\eqrefplain{eq:1}) using the loading matrix $\tilde{\Lambda}_{i}$:
\begin{equation}
\boldsymbol{z}_{i,j} = \beta_i(\boldsymbol{x}_{i,j}-\boldsymbol{\mu_i}) \label{eq:4}
\end{equation}
where $\boldsymbol{z}_{i,j} \in \mathbb{R}^{m}$, $\boldsymbol{x}_{i,j} \in \mathbb{R}^{n_i}$, and $\beta_i=\tilde{\Lambda}_{i}^\intercal(\tilde{\Lambda}_{i} \tilde{\Lambda}_{i}^\intercal+\Psi_i)^{-1}$. By using $\tilde{\Lambda}_{i}$, the induced latent activity $\boldsymbol{z}_{i,j}$ resides in a common latent space across all sessions.

\subsection{Stimulation-response model fitting} \label{model_fitting_method}
Using the merged neural activity across sessions, OMiSO fits statistical models (termed stimulation-response models, \cref{model_architecture_supfig}) to predict post-stimulation latent neural activity states. To train the stimulation-response models, OMiSO uses stimulation trials of $R$ experimental sessions collected in Section~\ref{alignment_method}, which involves stimulation with randomly chosen patterns among all possible stimulation patterns defined by the user. OMiSO creates datasets that comprise the pre-stimulation and post-stimulation latent activity states (where the latent space is defined using latent space alignment, Section~\ref{alignment_method}), and the stimulation patterns tested across the $R$ sessions. For the $k$th trial within a given session $i$, the pre-stimulation latent activity state over $t$ time bins $\boldsymbol{Z}^\text{Pre}_{i,k} \in \mathbb{R}^{m \times t}$ and induced post-stimulation latent activity state $\boldsymbol{z}^\text{Post}_{i,k} \in \mathbb{R}^{m}$ are computed using \eqrefplain{eq:4}. The entries of $\boldsymbol{Z}^\text{Pre}_{i,k}$ and $\boldsymbol{z}^\text{Post}_{i,k}$ corresponding to the user-defined $l$ target dimensions within the $m$ dimensional aligned latent space ($l\leq m$) are then subselected and collected in new vectors $\check{\boldsymbol{Z}}^\text{Pre}_{i,k} \in \mathbb{R}^{l \times t}$ and  $\check{\boldsymbol{z}}^\text{Post}_{i,k} \in \mathbb{R}^{l}$. Given a planar grid of electrodes of size $h \times v$ ($h, v \in \mathbb{R}$), the stimulation pattern tested in the $k$th trial during the $i$th session $S_{i,k}, \in \mathbb{R}^{h \times v}$ is encoded using a value of 1 for stimulated electrodes and 0 for all other electrodes. To train the stimulation-response model $f(\hat{\boldsymbol{z}}^\text{Post}_{i,k}|S_{i,k}, \check{\boldsymbol{Z}}^\text{Pre}_{i,k})$, which maps the stimulation pattern $S_{i,k}$ and pre-stimulation state $\check{\boldsymbol{Z}}^\text{Pre}_{i,k}$ to the predicted post-stimulation state $\hat{\boldsymbol{z}}^\text{Post}_{i,k} \in \mathbb{R}^{l}$, OMiSO minimizes the mean squared error (MSE) between the predicted $\hat{\boldsymbol{z}}^\text{Post}_{i,k}$ and actual $\check{\boldsymbol{z}}^\text{Post}_{i,k}$ post-stimulation latent states. OMiSO performs $R$-fold cross validation where every session is used as a validation set once (see Section~\ref{model_fitting_sup_method}). The final prediction $\hat{\boldsymbol{z}}^\text{Post}_{i,k}$ is obtained by averaging the predictions of the $R$ models.

\subsection{Inverting the stimulation-response models} \label{bc_method}
Using the trained stimulation-response models, OMiSO estimates a stimulation-response inverse model. The goal of the stimulation-response inverse model is to identify a stimulation pattern that can reach a target neural activity state given the pre-stimulation neural activity state. More specifically, a stimulation-response inverse model $\pi_{\theta}(\hat{\boldsymbol{\boldsymbol{s}}}_{i,k} | \check{\boldsymbol{z}}^\text{Targ}, \check{\boldsymbol{Z}}^\text{Pre}_{i,k})$, parameterized by $\theta$, receives the latent target state $\check{\boldsymbol{z}}^\text{Targ} \in \mathbb{R}^l$ and pre-stimulation state $\check{\boldsymbol{Z}}^\text{Pre}_{i,k} \in \mathbb{R}^{l \times t}$. Then, it returns a $u$-dimensional vector of stimulation suitabilities for each electrode $\hat{\boldsymbol{s}}_{i,k} \in \mathbb{R}^{u}$, where $u$ is the total number of candidate electrodes to be stimulated whose value is determined by the user and $u \le h \times v$. Each entry in $\hat{\boldsymbol{s}}_{i,k}$ is a number between 0 and 1 representing the suitability
of stimulating each electrode for achieving the latent target state, where a value closer to 1 indicates greater suitability. Note that the vector as a whole is not normalized to sum to one.

In this study, OMiSO uses Behavioral Cloning (BC) \citep{pomerleau1988alvinn, bain1995framework} to train a stimulation-response inverse model $\pi_\theta$ (\cref{model_architecture_supfig}). BC is a widely used approach in reinforcement learning (RL) to learn the optimal action given different states when expert data are readily accessible. An action in our case is to select a stimulation pattern. The state is a combination of a pre-stimulation latent state and a target latent state. OMiSO generates expert data using the trained stimulation-response models. These models output the expected average post-stimulation state over $R$ models $\hat{\boldsymbol{z}}^\text{Post} \in \mathbb{R}^{l}$ for any given combination of a stimulation pattern $S \in \mathbb{R}^{h \times v}$ and pre-stimulation state $\check{\boldsymbol{Z}}^\text{Pre} \in \mathbb{R}^{l \times t}$. By considering the predicted post-stimulation state $\hat{\boldsymbol{z}}^\text{Post}$ as a target state $\check{\boldsymbol{z}}^\text{Targ}$, OMiSO constructs expert data $E \in \{\check{\boldsymbol{Z}}^\text{Pre}, \check{\boldsymbol{z}}^\text{Targ}, \boldsymbol{s}\}$ where $\boldsymbol{s} \in \mathbb{R}^{u}$ is a vector representation of $S$ including only the $u$ candidate electrodes for stimulation. The value for each candidate electrode in $\boldsymbol{s}$ is 1 if stimulated and 0 if not. Expert data is used to train the inverse model so that it outputs the stimulation pattern needed to create a target latent state, given a pre-stimulation latent state. OMiSO trains the stimulation-response inverse model by minimizing the binary cross entropy over multiple epochs with respect to $\theta$ (Section~\ref{bc_sup_method}).

\subsection{Stimulation pattern selection on each trial} \label{pattern_selection_method}
To find the stimulation parameters
for achieving a target latent state with the stimulation-response inverse model, each brain stimulation session starts with latent space identification trials, where no stimulation is applied. OMiSO extracts a list of usable electrodes $\boldsymbol{e_c}$, which includes $n_c$ electrodes, using these trials. The observed spike count vectors $\boldsymbol{x}_{c,j} \in \mathbb{R}^{n_c}$ for all time bins $j$ across all these trials are used to fit the FA parameters $\Lambda_{c} \in \mathbb{R}^{n_c \times m}$, $\boldsymbol{\mu}_{c} \in \mathbb{R}^{n_c}$, and $\Psi_{c} \in \mathbb{R}^{n_c \times n_c}$. The identified latent space $\Lambda_{c}$ is aligned to the reference latent space $\Lambda_0$ using the methods described in Section~\ref{alignment_method}, yielding  $\tilde{\Lambda}_{c} \in \mathbb{R}^{n_c \times m}$. To start the optimization, OMiSO loads the user defined target state $\check{\boldsymbol{z}}^\text{Targ} \in \mathbb{R}^l$, the trained stimulation-response inverse model $\pi_\theta$, and the number of electrodes used for each stimulation pattern $n_\text{Stim} \in \mathbb{R}$. Note that out of $h \times v$ total electrodes (used for the stimulation pattern representation $S$), not all may be suitable for stimulation (e.g., some may have no effect on the neural population activity, Section~\ref{experimentprotocol_sup_method}). This results in $u$ ($\le h \times v$) candidate electrodes (used for the stimulation pattern representation $\boldsymbol{s}$). Of those, OMiSO chooses $n_\text{Stim}$ electrodes to form each stimulation pattern, where $n_\text{Stim}\le u$.

On each trial, OMiSO selects the stimulation pattern to perform given the observed pre-stimulation neural activity states using the inverse model $\pi_\theta$. On the $k$th trial, OMiSO estimates the pre-stimulation latent activity state $\boldsymbol{Z}^\text{Pre}_{c,k} \in \mathbb{R}^{m \times t}$ in real time using \eqrefplain{eq:4}. 
The entries of $\boldsymbol{Z}^\text{Pre}_{c,k}$ corresponding to the target dimensions are subselected as $\check{\boldsymbol{Z}}^\text{Pre}_{c,k} \in \mathbb{R}^{l \times t}$, and $\check{\boldsymbol{Z}}^\text{Pre}_{c,k}$ and $\check{\boldsymbol{z}}^\text{Targ}$ are passed to $\pi_\theta$ to get a $u$-dimensional vector of stimulation suitabilities for each candidate electrode $\hat{\boldsymbol{s}}_{c,k}$. Using $\hat{\boldsymbol{s}}_{c,k}$, OMiSO chooses a stimulation pattern with an epsilon greedy algorithm. With probability $1-\varepsilon_{c,k}$, where $0\le\varepsilon_{c,k}\le1$, it chooses the $n_\text{Stim}$ electrodes with the highest predicted suitabilities in $\hat{\boldsymbol{s}}_{c,k}$. With probability $\varepsilon_{c,k}$, it chooses $n_\text{Stim}$ electrodes stochastically following the softmax transformed probability $p^w_{c,k} \in \mathbb{R}$ computed for each $w$th electrode ($w=1,...,u$) as:
\begin{equation}
p^w_{c,k} = \frac{\exp(\hat{s}^w_{c,k})}{\sum_{v=1}^{u} \exp(\hat{s}^v_{c,k})}
\end{equation}
where $\hat{s}^w_{c,k} \in \mathbb{R}$ is the predicted stimulation suitability for the $w$th electrode. In this way, electrodes with low predicted stimulation suitability can be occasionally chosen to induce exploration, but the electrodes with higher stimulation suitabilities are chosen more often. 
OMiSO uses a time-varying $\varepsilon_{c,k}$ to initially encourage exploration and gradually shift to exploitation:  
\begin{equation}
\varepsilon_{c,k} = \max(\varepsilon_{\text{init}} \cdot \gamma^{k},~\varepsilon_{\text{floor}})
\end{equation}
where $0 \le\varepsilon_\text{init} \le 1$ is an initial value of $\varepsilon_{c,k}$, $0 \le \gamma \le 1$ is a discount factor, $0 \le\varepsilon_\text{floor} \le 1$ is a floor value of $\varepsilon_{c,k}$ ($\varepsilon_\text{floor}\le\varepsilon_\text{init}$), and $ k $ is the current trial index. OMiSO uses $\varepsilon_\text{floor}$ to maintain some amount of exploration throughout a session. Once the stimulation pattern is selected with the epsilon greedy algorithm, OMiSO performs stimulation with the selected $n_\text{Stim}$ electrodes and measures post-stimulation spike counts $\boldsymbol{x}^\text{Post}_{c,k} \in \mathbb{R}^{n_c}$ to compute $\check{\boldsymbol{z}}^\text{Post}_{c,k} \in \mathbb{R}^{l}$ (\eqrefplain{eq:4}). The set of $\check{\boldsymbol{z}}^\text{Targ}$, $\check{\boldsymbol{Z}}^\text{Pre}_{c,k}$, $\check{\boldsymbol{z}}^\text{Post}_{c,k}$, and the tested stimulation pattern $\boldsymbol{s}_{c,k} \in\mathbb{R}^u$ are stored to update the stimulation-response inverse model in a batched manner, as explained in the next section.

\subsection{Adaptive updating of the stimulation-response inverse model} \label{ppo_method}
During the brain stimulation experiment, OMiSO adaptively updates the stimulation-response inverse model $\pi_\theta$ using the stored observations. In this study, we implemented OMiSO's adaptive model update using a clipped policy gradient objective function used in Proximal Policy Optimization (PPO) \citep{schulman2017proximal}. PPO is a widely-used RL method to update an action selection strategy. In our case, PPO is used to update the stimulation-response inverse model to improve the stimulation pattern selection strategy. Compared to other policy update algorithms such as REINFORCE \citep{williams1992simple} and Actor Critic \citep{konda1999actor}, PPO offers a more stable policy update by using the clipped objective function, which might be particularly well suited to handle large trial-by-trial variance in neural activity. 

At each update, OMiSO increases the suitability of stimulating electrodes that produced a post-stimulation state closer to the target state and decreases it for electrodes that did not. Specifically, for $k$th stored trial, OMiSO computes the advantage $A_{c,k} \in \mathbb{R}$ for the stimulation pattern used in the trial:
\begin{equation}
\begin{matrix}
A_{c,k} = -||\check{\boldsymbol{z}}^\text{Targ}-\check{\boldsymbol{z}}^\text{Post}_{c,k}||_1+b
\label{eq:6}
\end{matrix}
\end{equation}
where $b \in \mathbb{R}$ is a fixed baseline value to evaluate the induced post-uStim activity.
The advantage $A_{c,k}$ reflects the effectiveness of the stimulation pattern $\boldsymbol{s}_{c,k}$ applied on the $k$th trial. If the pattern achieves an absolute error smaller than the baseline, the advantage is positive; otherwise, it is negative. With the computed advantage $A_{c,k}$, OMiSO updates the parameters $\theta$ in the stimulation-response inverse model $\pi_\theta$ by performing gradient ascent over multiple epochs on the following objective function:
\begin{equation}
\begin{matrix}
\mathcal{L}_\text{PPO} = \sum_{e=1}^{n_\text{Stim}} \min \left(r_{c,k}^{e}(\theta) A_{c,k}, \operatorname{clip} \left( r_{c,k}^{e}(\theta), 1 - \varepsilon_\text{clip}, 1 + \varepsilon_\text{clip} \right) A_{c,k} \right), \\
\text{where}\hspace{3mm} r_{c,k}^{e}(\theta) = \frac{\pi_{\theta}(\hat{s}_{c,k}^{e} | \check{\boldsymbol{z}}^\text{Targ}, \check{\boldsymbol{Z}}^\text{Pre}_{c,k})}{\pi_{\theta_{\text{old}}}(\hat{s}_{c,k}^{e} | \check{\boldsymbol{z}}^\text{Targ}, \check{\boldsymbol{Z}}^\text{Pre}_{c,k})},\\
\label{eq:8}
\end{matrix}
\end{equation}
$e=1,...,n_\text{Stim}$ is the index of the electrode among $n_\text{Stim}$ electrodes used for the stimulation pattern $\boldsymbol{s}_{c,k}$, and $0\le\varepsilon_\text{clip}\le1$ helps to limit the range of the policy gradient update to avoid taking too large of an update step
(Section~\ref{parameters_sup_method}). $\pi_{\theta}(\hat{s}_{c,k}^{e} | \check{\boldsymbol{z}}^\text{Targ}, \check{\boldsymbol{Z}}^\text{Pre}_{c,k})$ represents the predicted stimulation suitability for the $e$th stimulated electrode under the updated stimulation-response inverse model. Similarly, $\pi_{\theta_{\text{old}}}(\hat{s}_{c,k}^{e} | \check{\boldsymbol{z}}^\text{Targ}, \check{\boldsymbol{Z}}^\text{Pre}_{c,k})$ is the predicted suitability under the original stimulation-response inverse model before updating $\theta$. The ratio $r_{c,k}^{e}(\theta)$ is the change in suitability for selecting the $e$th electrode. To maximize this objective function, PPO updates $\theta$ to increase $r_{c,k}^{e}(\theta)$ when the advantage is positive and decrease it when the advantage is negative.

\subsection{Experiment details} \label{experiment_method}
We tested OMiSO using uStim in a macaque monkey with a 96-electrode Utah array implanted in the PFC. In each experimental session, the monkey performed a memory-guided saccade task (Section~\ref{experimentprotocol_sup_method}). Each trial began when the monkey fixated a central dot, followed by a memory target briefly appearing at a peripheral location (Section~\ref{experimentprotocol_sup_method}). After the memory target disappeared, the monkey needed to remember the target location during a "memory period" (\cref{overview_fig}B). The go cue was signaled by the disappearance of the central dot,
after which the monkey saccaded to the remembered location. On about 80\% of trials, we applied uStim for 40 ms during the memory period (\cref{overview_fig}B). The stimulation patterns tested in this study consisted of all possible $n_\text{Stim}$=5 electrode spatial patterns chosen from $u$=20 candidate electrodes (15,504 total patterns, \cref{candidate_elec_supfig}, see Section~\ref{experimentprotocol_sup_method} for the selection criteria of the candidate electrodes). 
Each experimental session started with 120 no-uStim trials for latent space identification. We identified $m$=5 latent dimensions by applying FA to spike counts taken in 50 ms bins. Among them, we chose $l$=2 target dimensions (Section~\ref{experimentprotocol_sup_method}). Post-uStim spike counts (50 ms bin) were calculated starting 10 ms after uStim offset (\cref{overview_fig}B, pink bar) to avoid uStim artifacts. The number of sessions used in each analysis is summarized in Section~\ref{numsession_sup_method}. 

\section{Results}

\begin{figure}[!tb]
\centering
\includegraphics[scale=0.85]{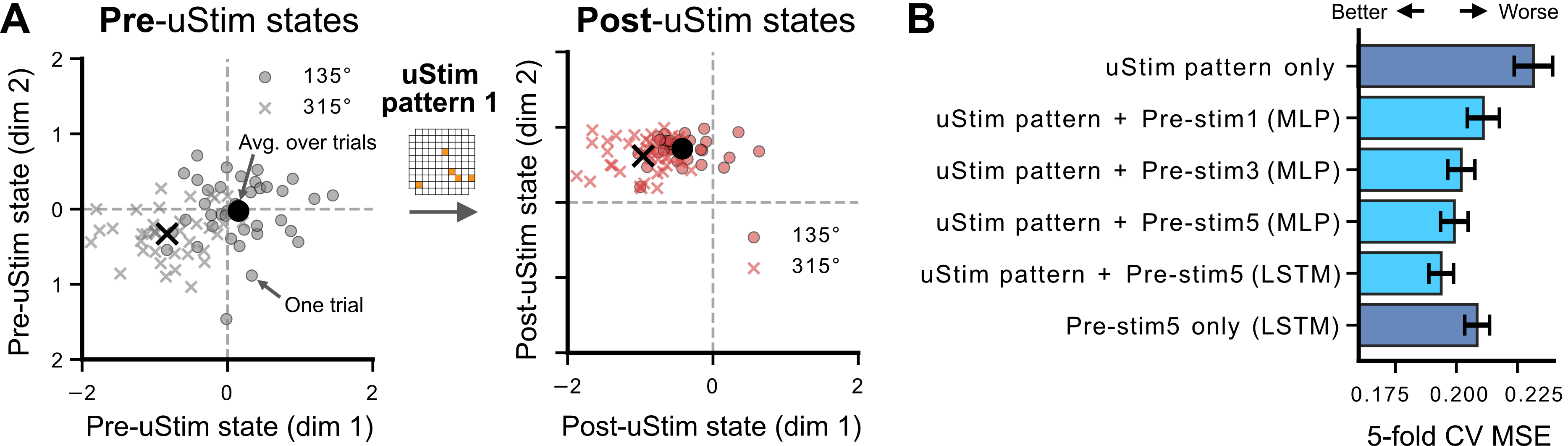}
\caption{\textbf{Pre-uStim brain states affect post-uStim brain states.} (A) Example 5-electrode uStim pattern. The orange cells in the array map indicate the location of the stimulated electrodes. Each small circle (135° memory condition) and cross (315° memory condition) indicate the neural population activity state at each single trial, and the large black circle and cross indicate the mean state across trials within each memory condition. (B) Post-uStim state prediction performance across models. Error bars indicate standard error across cross-validation folds (\cref{model_fitting_method}).}
\label{pre_vs_post_fig}
\end{figure}

\subsection{Brain's response to stimulation depends on pre-uStim brain state} \label{single_trial_var_result}
Does neural population activity prior to brain stimulation affect the brain's response to that stimulation? To assess this, we measured the impact of pre-uStim state on uStim response in a behavioral experiment where different memory conditions (see Section~\ref{experimentprotocol_sup_method}) were used to create distinct pre-uStim neural population activity states in low-d space (\cref{pre_vs_post_fig}A, left). For all tested uStim patterns, uStim shifted neural activity within the low-d space (\cref{pre_vs_post_fig}A, right, \cref{pre_vs_post_supfig}) away from the location where the activity would be without uStim (\cref{pre_vs_post_supfig}, leftmost column, bottom panel). This demonstrates that uStim perturbed the neural population activity. The induced post-uStim activity states were separable between two memory conditions (\cref{pre_vs_post_fig}A, right, \cref{pre_vs_post_supfig}). These results suggest that neural population activity states induced by uStim depend on the pre-uStim neural activity state.

To confirm whether this state dependency helps improve the prediction of brain response to uStim, we compared the post-uStim state prediction performance of statistical models with and without pre-uStim state information (\cref{model_architecture_supfig}). For a model without pre-uStim state information, we used a CNN which predicted post-uStim states only as a function of the uStim patterns applied (as in \citep{minai2024miso}), termed ``uStim pattern only''. For models with pre-uStim state information, we tested a combination of a CNN and a Multi Layered Perceptron (MLP) \citep{rumelhart1986learning} with up to 5 pre-uStim time bins, termed ``uStim pattern + Pre-stim\{1,3,5\} (MLP)'', and a combination of a CNN and an LSTM with 5 pre-uStim time bins, termed ``uStim pattern + Pre-stim5 (LSTM)''. To collect training data to fit these models, we applied randomly-chosen 5-electrode uStim patterns in 6 experimental sessions with a monkey.

The models with pre-uStim state information (\cref{pre_vs_post_fig}B, four middle bars in light blue) outperformed the model without this information (\cref{pre_vs_post_fig}B, top dark blue bar, $p$<0.05 for all four models, one-tailed t-test). Among the models with pre-uStim state information, uStim pattern + Pre-stim5 (LSTM) achieved the best prediction performance by incorporating 5 time bins of pre-uStim state information. A model using only pre-uStim state information, termed ``Pre-stim5 only (LSTM)'' performed worse than the uStim pattern + Pre-stim5 (LSTM) model (\cref{pre_vs_post_fig}B, bottom dark blue bar, $p$<0.05, one-tailed t-test). Thus, both uStim pattern and pre-uStim state information were important for prediction. These results indicate that the pre-uStim state affects where the activity evolves after uStim, and that this can be leveraged to improve the prediction of brain response to uStim.

\begin{figure}[!tb]
\centering
\includegraphics[scale=0.9]{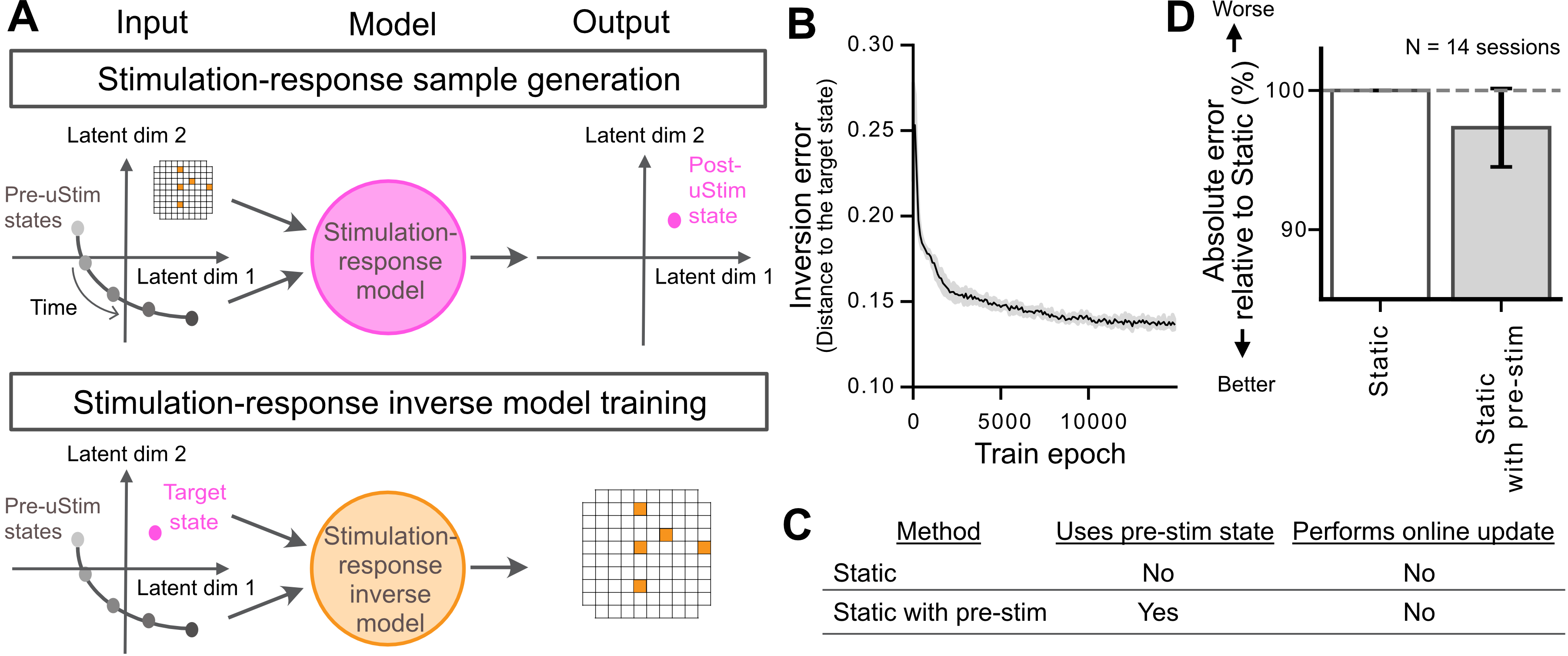}
\caption{\textbf{Stimulation-response model inversion to choose a uStim pattern based on pre-uStim state.} (A) Schematic of the inversion procedure. (B) Inversion performance. The error bar indicates the standard deviation across 5 inverse models. (C) Methods used for performance comparison. (D) Absolute error relative to the ``Static'' method computed in a 2-d target latent space. Error bars indicate standard error across sessions.}
\label{bc_fig}
\end{figure}

\subsection{A stimulation-response inverse model selects uStim patterns based on pre-uStim states} \label{behavior_cloning_result}
OMiSO's goal is to choose a uStim pattern that can create a targeted neural population state. While the stimulation-response models trained in Section~\ref{single_trial_var_result} can predict brain responses to any input combination of a uStim pattern and a pre-uStim state (\cref{bc_fig}A, top), they cannot directly generate a uStim pattern for a target brain state. One potential approach is to construct a table of predicted post-uStim states for all possible input combinations and find the uStim pattern expected to produce the closest post-uStim state to the target state. However, this approach is infeasible with a continuum of pre-uStim states. Another approach is to mathematically invert the stimulation-response models to choose a uStim pattern. However, this is challenging because the models could map multiple combinations of uStim pattern and pre-uStim state to the same post-uStim state (i.e., the mapping is not one-to-one).

Instead, we inverted the stimulation-response models using behavioral cloning (BC) (Section~\ref{bc_method}). The inverse model received a pre-uStim state and a target activity state and returned a suitability of stimulating each candidate electrode (\cref{bc_fig}A, bottom). To train the inverse models, we generated expert data using the stimulation-response models, where we considered the model-predicted post-uStim states as potential target states. For the stimulation-response model, we used the CNNxLSTM model with 5 pre-uStim states, which achieved the best prediction performance (\cref{pre_vs_post_fig}C). For the stimulation-response inverse model, we used an MLP (\cref{model_architecture_supfig}), since it is computationally fast and could complete all necessary computations within 50 ms (\cref{overview_fig}B, computation period) following the observation of pre-uStim states, allowing uStim to be delivered immediately afterward.

To evaluate inversion quality, we checked if the inverse model could output uStim patterns similar to those used by the stimulation-response model to generate post-uStim state predictions. We found that with more epochs, the inverse model got better at choosing uStim patterns that induced post-uStim states close to the original predictions (\cref{bc_fig}B, Manhattan distance in 2-d latent space, \eqrefplain{eq:6}). This improvement saturated around 100,000 epochs with a distance (the inversion error) of around 0.13 a.u. (error for perfect inversion is 0). For comparison, the trial-by-trial variance in neural activity states (i.e., the std. of latent states in no-uStim trials within the same memory condition) is 0.7 a.u.

\begin{figure}[!tb]
\centering
\includegraphics[scale=0.85]{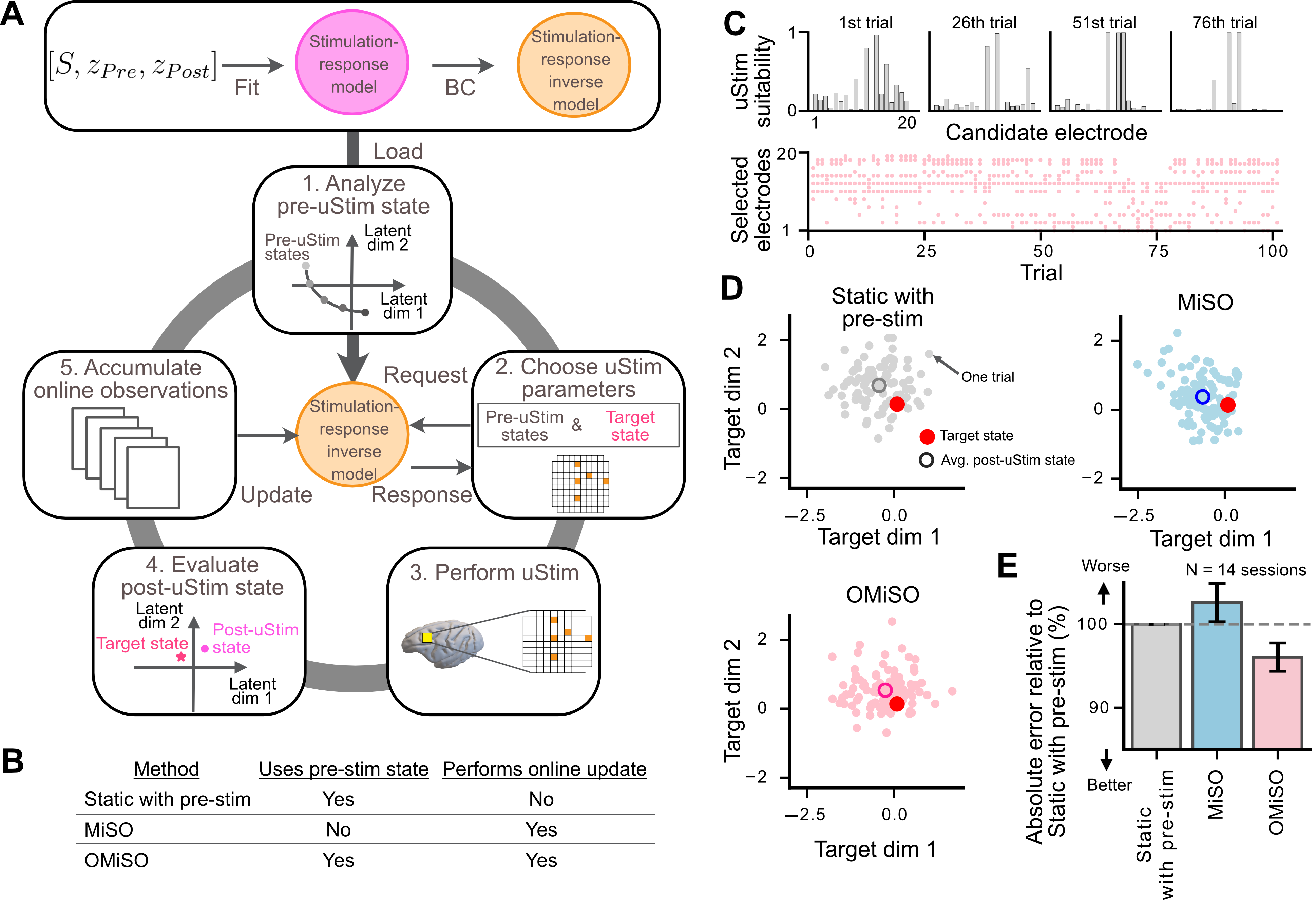}
\caption{\textbf{Performance of OMiSO in monkey experiments.} (A) Schematic of OMiSO. OMiSO trains the stimulation-response model and a stimulation-response inverse model using previously collected neural activity (top box). In each stimulation trial, OMiSO performs five steps using the trained inverse model. (B) Methods used for performance comparison. (C) Examples of stimulation suitability for each electrode (top) and selected uStim patterns (bottom) by OMiSO during an example session. (D) Induced post-uStim states by each method during an example session (the same session as in panel C). (E) Absolute error of different methods relative to ``Static with pre-stim'' computed in 2-d target latent space. Error bars indicate standard error across sessions.}
\label{ppo_fig}
\end{figure}

To understand the impact of imperfect model inversion on achieving target brain states, we experimentally compared the performance of two methods, one that could be perfectly inverted and another that could only be approximately inverted. The first method constructed a table of predicted post-uStim states for all possible uStim patterns (15,504 patterns) using ``uStim pattern only'' model, which does not consider pre-uStim state information. It then found the uStim pattern expected to produce the closest post-uStim state to the target state. Access to the table implies a \textit{perfect} inversion of the model. The second method used the inverse model of the CNNxLSTM model with 5 pre-uStim states trained with BC (which could only achieve an \textit{imperfect} inversion). Neither the prediction table nor the inverse model were updated during the experiments, so we refer to them as the ``Static'' and ``Static with pre-stim'' methods, respectively (\cref{bc_fig}C). To compare performance, we asked each method to choose and perform uStim patterns to create a 2-d target neural activity state. If the effect of the inversion error on the uStim pattern selection performance is small, the ``Static with pre-stim'' model, which leveraged pre-uStim information, should outperform the ``Static'' method. We found that the mean error to the target state by ``Static with pre-stim'' was smaller than for the ``Static'' method, although the difference was not statistically significant (\cref{bc_fig}D, $N$=14 sessions, $p$=0.3, Wilcoxon signed-rank test). This result suggests that the inverse model trained with BC may be successfully incorporated into an adaptive brain stimulation framework to choose uStim patterns based on pre-uStim state information.

\subsection{OMiSO adaptively updates the inverse model during a brain stimulation experiment} \label{ppo_result}
During each experimental session, new stimulation-response samples become available. Leveraging these samples may help us adaptively update the stimulation-response inverse model to account for inversion errors, as well as changes in the stimulation-response relationship within and across sessions. To test this, we developed an adaptive brain stimulation framework, OMiSO (\cref{ppo_fig}A, Section~\ref{generalframework}) where we used PPO to adaptively update the stimulation-response inverse model (Section~\ref{ppo_method}). We tested OMiSO during brain stimulation experiments with a monkey and compared its performance against an inverse model that was not adaptively updated (``Static with pre-stim'', tested in Section~\ref{behavior_cloning_result}). OMiSO had limited observations to update the inverse model, which could put it at a disadvantage due to overfitting compared to the ``Static with pre-stim'' model. We also compared OMiSO's performance with a previously proposed method, MiSO \citep{minai2024miso}, which updates its predictions but does not leverage pre-uStim states (\cref{ppo_fig}B). OMiSO adaptively updated the stimulation suitability for each electrode (\cref{ppo_fig}C, top; example session), changing the uStim pattern applied (\cref{ppo_fig}C, bottom), which allowed it to drive the post-uStim states closer to the target state than the other two methods (\cref{ppo_fig}D). Across multiple sessions, OMiSO achieved significantly smaller errors to the target state than both alternative methods (\cref{ppo_fig}E, $N$=14 sessions, $p$=0.05 for Static with pre-stim and $p$=0.01 for MiSO, Wilcoxon signed-rank test) by adaptively updating the stimulation-response inverse model and taking pre-uStim state information into account.

\section{Discussion}
\label{disc}
We propose a brain stimulation framework OMiSO, which involves two key methodological advances for shaping neural population activity states: incorporation of the pre-stimulation brain state and adaptive updating of the model used to choose stimulation parameters. In monkey experiments, OMiSO outperformed methods that do not incorporate these advances. While the neural activity used in this study consists of spiking responses recorded with a planar grid of electrodes implanted in the brain, and each stimulation pattern specifies the location of the stimulated electrodes within the grid, OMiSO can be readily applied to other stimulation and recording protocols (e.g., holographic optogenetics \citep{adesnik2021probing} or other uStim parameters).

Although we experimentally demonstrated the advantages of the incorporation of the pre-stimulation brain state, there are scenarios where it might be less beneficial. For example, in response to strong stimulation or if the neural activity is governed by attractor dynamics, the post-stimulation activity might be driven to a similar state regardless of the pre-stimulation activity. Another scenario is when there is a significant shift in the distribution of neural states between the training period and ongoing use. In clinical applications, the neural states can differ substantially between healthy and disease conditions. For example, individuals with epilepsy exhibit bursting activity, which is not usually observed under healthy conditions \citep{thijs2019epilepsy}. Before using OMiSO, one should assess the alignment of the neural state distributions between the training period and ongoing use. 

OMiSO's adaptive model update uses the newly observed stimulation-response samples within the same session (i.e., on-policy RL).  One could consider also using the samples observed in previous sessions to improve the sample efficiency (i.e., off-policy RL), possibly using a replay buffer \citep{mnih2013playing}. With these advances, OMiSO provides greater accuracy in achieving targeted activity states over longer periods of time, thereby improving neuromodulation technologies for understanding the brain and for treating brain disorders.

\section{Acknowledgments}
This work was supported by Japan Student Services Organization fellowship, NSF NCS DRL 2124066, and Simons Foundation NC-GB-CULM-00003241-05. The authors are inventors on pending US Patent Application No. 63/697,896 and 63/795,999, which relate to the adaptive stimulation methods developed in this paper.

\newpage

\bibliographystyle{unsrt}
\bibliography{reference}

\clearpage
\appendix

\renewcommand{\thefigure}{S\arabic{figure}}
\setcounter{figure}{0}
\renewcommand{\thesection}{S\arabic{section}}
\setcounter{section}{0}

\textbf{\Large Supplementary material}

\section{Summary of hyperparameters} \label{parameters_sup_method}
The table below summarizes the hyperparameter values used in this study. The latent dimensionality for the FA model $m$ was chosen by first, computing the optimal dimensionality separately for each session based on cross-validated data likelihoods, and second, finding the most frequently observed optimal dimensionality across all sessions. The number of stable electrodes for alignment was determined by assessing the number of usable electrodes with consistent firing activity patterns across sessions during the memory period. The learning rate, weight decay, and training epoch for stimulation-response model training were chosen based on a grid search. We did not perform extensive hyperparameter tuning for the parameters for BC, since it was robust to parameter choices. The parameters for uStim pattern selection and PPO were chosen by running simulations (Section~\ref{ppo_hyperparameter_tuning_sup_method}). uStim parameters (amplitude, frequency, and duration) were set not to cause overt behavioral changes or strong post-uStim activity inhibition across the whole array, both of which can occur in some uStim parameter regimes.

\begin{table}[htbp]
  \centering
  \resizebox{\textwidth}{!}{%
\begin{tabular}{lc c c c cl}\toprule
            Method & Parameter & Description & Value \\\midrule
\midrule
\ Latent space  & $m$ & FA latent dimensionality & 5 \\
\ identification \& alignment & $n_\text{stable}$ & Num. of stable electrodes for alignment & 40 \\\midrule
\ Stimulation-response & $R$ & Num. of training experimental sessions & 5 \\
\ model fitting & $t$ & Num. of pre-uStim state time bins & 1-5 \\  & $l$ & Target state dimensionality & 2 \\
\   & - & Optimizer & AdamW \\
\  & - & Learning rate & 0.0006 \\
\  & - & Weight decay & 0.001 \\
\  & - & Training epoch & 20
\\\midrule
\ Behavioral Cloning & - & Batch size per epoch & 15,504 \\
\  & - & Max epoch & 100,000\\
\  & - & Num. of validation samples & 1,000\\
\  & - & Validation frequency (epoch) & 100\\
\  & - & Early stop patience (epoch) & 20\\\midrule
\ uStim pattern selection & - & Num. of latent space identification trials & 120\\
\  & $\varepsilon_\text{init}$ & Initial value of $\varepsilon_{c,k}$ & 0.5\\
\  & $\varepsilon_\text{floor}$ & Minimum value of $\varepsilon_{c,k}$ & 0.1\\
\  & $\gamma$ & Discount factor of $\varepsilon_{c,k}$ & 0.95, 0.96\\\midrule
\ PPO & $\varepsilon_\text{clip}$ & Clip value of PPO update & 0.15, 0.2\\
\  & $b$ & Baseline for advantage computation & 0.6, 0.7 \\
\  & - & PPO update frequency (number of trials) & 5\\
\  & - & Optimizer & AdamW\\
\  & - & Learning rate & $2 \times 10^{-5}$, $1 \times 10^{-4}$\\
\  & - & Weight decay & 0.001\\\midrule
\ uStim configuration & $u$ & Num. of uStim candidate electrodes & 20 \\
\  & $n_\text{Stim}$ & Num. of uStim electrodes & 5 \\
\  & - & Total num. of possible uStim patterns & 15,504 \\
\  & $h$ & Horizontal size of electrode grid & 10 \\
\  & $v$ & Vertical size of electrode grid & 10 \\
\  & - & uStim amplitude (uA) & 25 \\
\  & - & uStim frequency (Hz) & 50 \\
\  & - & uStim duration (ms) & 40
\\\bottomrule
\end{tabular}
}
\end{table}

\section{Spiking activity preprocessing} \label{preprocessing_sup_method}
To identify latent dimensions of the neural population activity using FA, we computed binned spike counts during the 1.5~s memory period for each of the first 120 trials recorded in each session (termed the ``latent space identification trials''). We used 50 ms bins, yielding 30 bins per trial. The total number of time bins used to fit the FA model on each session was 3,600 (120 trials $\times$ 30 bins). The same trials were used to extract a list of usable electrodes $\boldsymbol{e_i}$ for each $i$th session based on the following three criteria: mean firing rate >1 Hz, Fano factor <8, and <20$\%$ coincident spiking with each of the other electrodes. The FA model was fitted using only the usable electrodes. 

\section{Details of uStim experimental paradigm} \label{experimentprotocol_sup_method}
Experimental procedures were approved by the Institutional Animal Care and Use Committee.
In each experimental session, the monkey performed a memory-guided saccade task. On each trial, the monkey first fixated on a dot at the center of the screen. After establishing fixation, a target appeared on the screen for a brief period of time (100 ms). This was followed by a memory period, after which the center dot turned off (go cue) and the monkey performed an eye movement to the remembered target location to receive a liquid reward. The total fixation duration was set at either 1.65 or 1.95 seconds for latent space identification trials and 1.25 or 1.55 seconds for other trials. The location of the target dot was chosen from four peripheral targets ([45°, 135°, 225°, 315°] or [0°, 70°, 135°, 270°]) for latent space identification and from two peripheral targets ([135°, 315°] or [0°, 70°]) for other trials. These target directions were chosen based on the mapped receptive fields of the recorded neurons so that diverse neural population activity states could be induced.

\begin{figure}[!tb]
\centering
\includegraphics[scale=1]{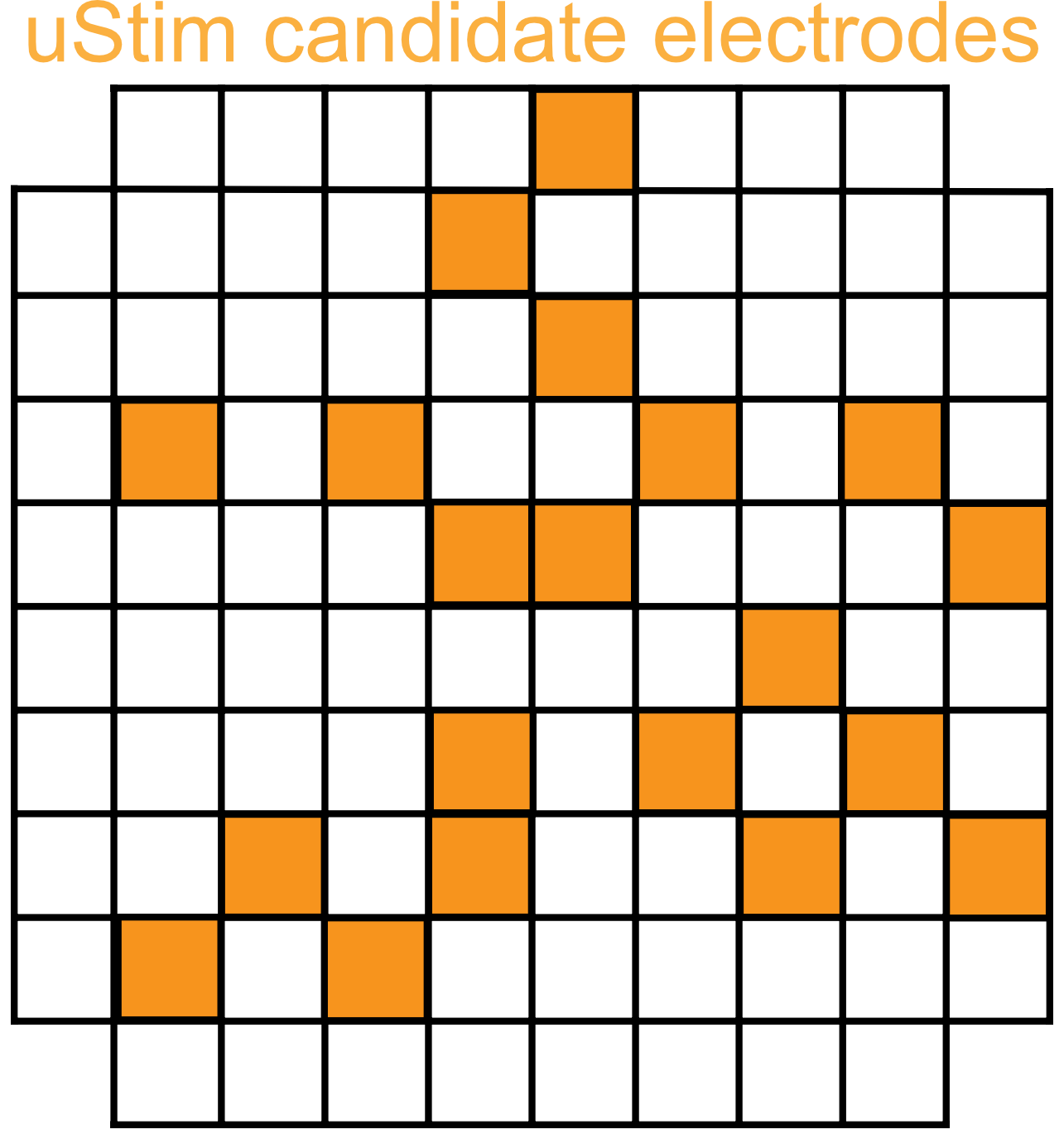}
\caption{\textbf{uStim candidate electrodes.} The orange cells indicate the location of the 20 candidate electrodes chosen based on the criteria described in Section~\ref{experimentprotocol_sup_method}.}
\label{candidate_elec_supfig}
\end{figure}

There were two types of memory-guided saccade trials: ``uStim trials'', in which we applied uStim, and ``no-uStim trials'', in which we did not apply uStim. The experimental system randomly chose which trial type to perform in an interleaved manner. On uStim trials, we applied uStim for 40 ms during the memory period. The stimulation was biphasic with each square pulse in the biphasic pair being 250 us in duration. We set the current amplitude low enough not to induce any eye movements (25 uA for each electrode we stimulated). On each trial, we changed the locations of $n_\text{Stim}$=5 stimulated electrodes selected among $u$=20 candidate electrodes (\cref{candidate_elec_supfig}), while keeping other parameter values such as current amplitude and frequency (50 Hz) fixed. The candidate electrodes for stimulation were selected based on three criteria: (1) they were spatially distributed across the array to prevent excessive current concentration when stimulating with multiple electrodes, (2) they were positioned near electrodes with consistent spiking activity across sessions to ensure reliable evaluation of brain responses to stimulation, and (3) they induced changes in brain activity when stimulated.

In each brain stimulation test session where we compared the performance of different methods, we chose two target latent dimensions satisfying two conditions: 1) they aligned well across multiple sessions, and 2) diverse latent activity states could be induced along these dimensions with uStim. The dimensions that passed the criteria were not necessarily the top FA dimensions that explained the greatest covariance among the neurons. In this study, to select which uStim pattern to apply, we chose the $n_\text{Stim}$ electrodes with the highest predicted suitabilities, with $1-\varepsilon_{c,k}$ probability (Section~\ref{pattern_selection_method}). We chose a fixed number of electrodes to stimulate ($n_\text{Stim}$) because we also stimulated with the same fixed number of electrodes during the training sessions (and we did so to simplify the training data collection process). However, OMiSO is flexible and could be adapted to other selection strategies. For example, one could apply a suitability threshold (e.g., 0.8) and stimulate all electrodes whose predicted suitability exceeds that threshold.

In our OMiSO implementation, the stimulation-response inverse model was adaptively updated every 5 trials using PPO at a different cluster machine than the machine used to run the experimental code. In this way, the model updates did not interfere with the stimulation parameter selection process, which is very time-sensitive. OMiSO checked the model file at the end of every trial and loaded the parameters of the model when this had been updated. For the PPO update, OMiSO only used new observations without using a replay buffer (i.e., it performed on-policy model updates), following Schulman et al. 2017's PPO implementation \citep{schulman2017proximal}. 

\section{Details of model fitting and computing resources} \label{model_fitting_sup_method}
The model architectures used in this study are depicted in \cref{model_architecture_supfig}.  Hyperparameters are listed in Section~\ref{parameters_sup_method}. All models were implemented in PyTorch \citep{paszke2019pytorch} and fit using the AdamW optimizer. We trained all models on a local computing cluster using 4 NVIDIA GeForce RTX GPUs and 11GB of RAM. The same local computing cluster was used to run PPO updates of the stimulation-response inverse model during uStim experiments. 

\section{Details of behavioral cloning} \label{bc_sup_method}
\begin{figure}[!tb]
\centering
\includegraphics[scale=1]{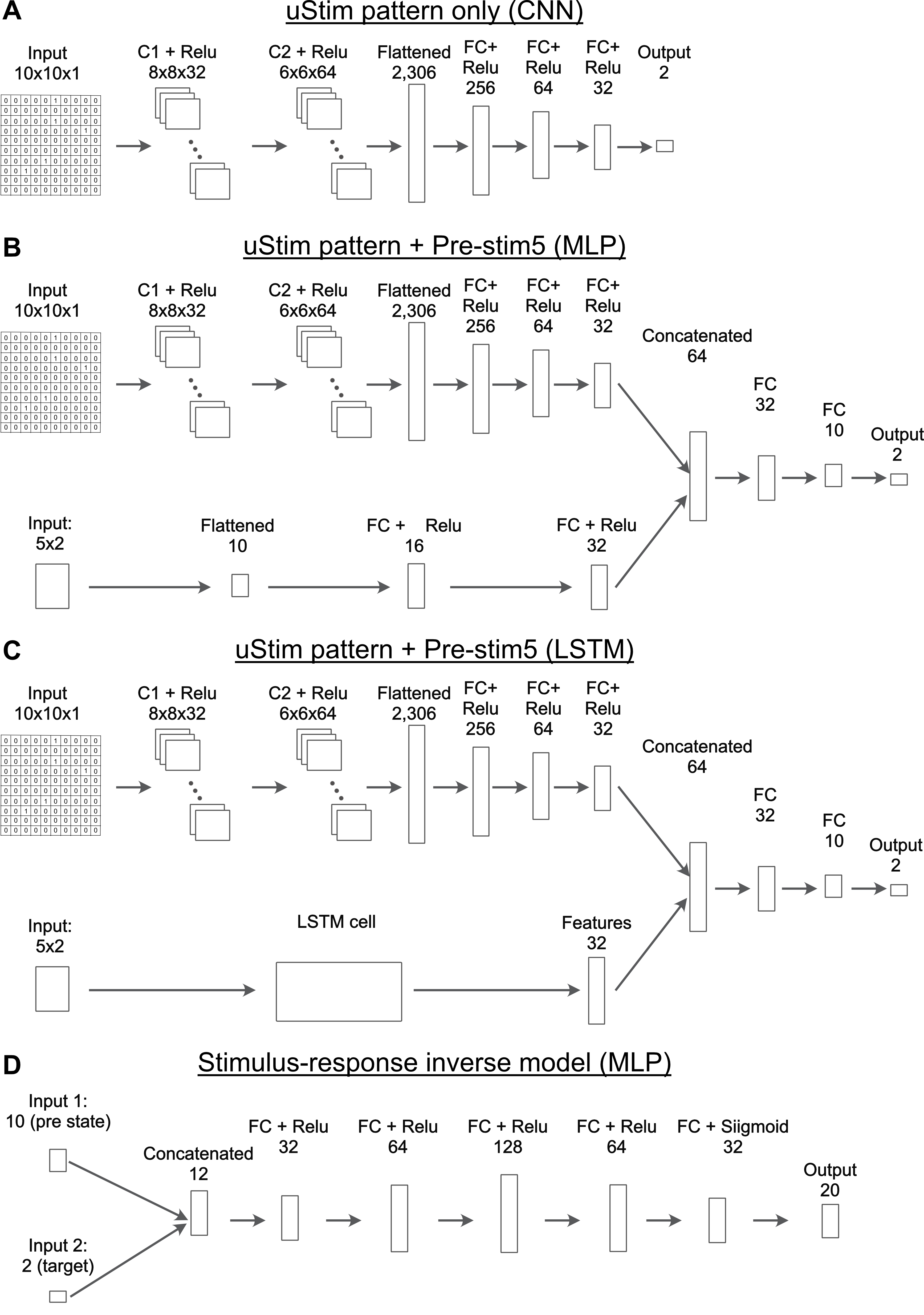}
\caption{\textbf{Model architectures.} (A) Stimulation-response model architecture, uStim pattern only \citep{minai2024miso} (\cref{pre_vs_post_fig}C). (B) Stimulation-response model architecture, uStim pattern + Pre-stim5 (MLP) (\cref{pre_vs_post_fig}C). (C) Stimulation-response model architecture, uStim pattern + Pre-stim5 (LSTM) (\cref{pre_vs_post_fig}C). (D) Stimulation-response inverse model architecture, MLP.}
\label{model_architecture_supfig}
\end{figure}

To train the stimulation-response inverse model (\cref{model_architecture_supfig}) with BC, OMiSO iterates the expert data generation process and model parameter updates over multiple epochs. On each epoch, OMiSO generates 15,504 expert samples, each of which consists of a pre-uStim state $\check{\boldsymbol{Z}}^\text{Pre} \in \mathbb{R}^{l \times t}$, a uStim pattern $\boldsymbol{s} \in \mathbb{R}^u$, and an average predicted post-uStim state across $R$ stimulation-response models, which is used as a target state $\check{\boldsymbol{z}}^\text{Targ}$. To generate one expert sample, OMiSO selects one of the experimentally observed pre-uStim states $\check{\boldsymbol{Z}}^\text{pre}_{i,k}$ as $\check{\boldsymbol{Z}}^\text{pre}$ and pairs it with a randomly chosen uStim pattern among 15,504 possible uStim patterns. OMiSO then uses the paired pre-uStim state and uStim pattern to generate the predicted target post-uStim state $\check{\boldsymbol{z}}^\text{Targ}$ using the stimulation-response models. With this expert data, OMiSO trains a stimulation-response inverse model by minimizing the binary cross entropy loss between the applied uStim patterns $\boldsymbol{s}$ and the predicted patterns $\hat{\boldsymbol{s}}$ with respect to $\theta$:
\begin{equation}
\begin{matrix}
\mathcal{L}_\text{BCE} = - \left[ \boldsymbol{s} \log (\hat{\boldsymbol{s}}) + (1 - \boldsymbol{s}) \log (1 - \hat{\boldsymbol{s}}) \right]\\
\hat{\boldsymbol{s}} = \pi_{\theta}(\check{\boldsymbol{z}}^\text{Targ}, \check{\boldsymbol{Z}}^\text{Pre})
\end{matrix} \label{eq:5}
\end{equation}
where $\hat{\boldsymbol{s}} \in \mathbb{R}^{u}$.

To evaluate the inversion quality, OMiSO uses 1,000 validation samples, which are separately generated using the stimulation-response models. Each $k$th validation sample consists of a pre-uStim state $\check{\boldsymbol{Z}}^\text{Pre}_{k} \in \mathbb{R}^{l \times t}$, a uStim pattern $\boldsymbol{s}_k \in \mathbb{R}^u$, and the target post-uStim state $\check{\boldsymbol{z}}^\text{Targ}_k \in \mathbb{R}^{l}$. During the validation, for each $k$th sample, OMiSO first obtains the predicted stimulation suitability of each candidate electrode using the inverse model:
\begin{equation}
\begin{matrix}
\hat{\boldsymbol{s}}_k = \pi_{\theta}(\check{\boldsymbol{z}}^\text{Targ}_k, \check{\boldsymbol{Z}}^\text{Pre}_k)
\end{matrix}
\end{equation}
It then chooses $n_\text{Stim}$ electrodes with the highest predicted suitabilities in $\hat{\boldsymbol{s}}_k$. Using a grid format representation of the selected uStim pattern $\hat{\boldsymbol{S}}_k \in \mathbb{R}^{h \times v}$ (the value is 1 for stimulated electrode locations and 0 for all other locations), OMiSO obtains the post-uStim state using the stimulation-response models:
\begin{equation}
\begin{matrix}
\hat{\boldsymbol{z}}^\text{Post}_{k} = f(\hat{\boldsymbol{S}}_k, \check{\boldsymbol{Z}}^\text{Pre}_k)
\end{matrix}
\end{equation}
Finally, OMiSO evaluates the error between the obtained post-uStim state and the target state across all validation samples:
\begin{equation}
\begin{matrix}
\mathcal{L}_\text{BC} = \frac{1}{1000} \sum_{k=1}^{1000} ||\check{\boldsymbol{z}}^\text{Targ}_k-\hat{\boldsymbol{z}}^\text{Post}_{k}||_1
\end{matrix} \label{eq:6}
\end{equation}
When the inversion with BC is perfect, this error becomes 0. OMiSO performs the validation every 100 epochs. To avoid overfitting, OMiSO stops the inverse model BC training when \eqrefplain{eq:6} does not improve for 20 consecutive validation performance evaluations.

\section{Summary of experimental sessions} \label{numsession_sup_method}
For this study, we collected 31 brain stimulation experimental sessions in total. Out of these sessions, 7 sessions were used for offline analyses (``Offline analysis sessions''), 10 sessions with randomly selected uStim patterns were used to train the stimulation-response models and the stimulation-response inverse model (``Training sessions''), and 14 sessions were used to test the performance of different methods (``Test sessions''). Among the 14 test sessions, 6 of them were run over the course of three days (two test sessions per day). The table below summarizes the number of sessions used for each result reported in the manuscript. 

\begin{table}[htbp]
  \centering
\begin{tabular}{lc c c c cl}\toprule
            Figure & Offline analysis sessions & Training sessions & Test sessions \\\midrule
\midrule
\ \cref{pre_vs_post_fig}B & 1 session & - & - \\
\ \cref{pre_vs_post_fig}C & 6 sessions & - & - \\
\ \cref{bc_fig}, \cref{ppo_fig} & - & 10 sessions & 14 session
\\\bottomrule
\end{tabular}
\end{table}

\section{Simulation to determine experiment hyperparameters} \label{ppo_hyperparameter_tuning_sup_method}
To decide the values of hyperparameters for adaptive brain stimulation experiments (listed in Section~\ref{parameters_sup_method}), we ran simulations by using the trained $R$ stimulation-response models (Section~\ref{model_fitting_method}) and the stimulation-response inverse model (Section~\ref{bc_method}). On each simulation, we simulated 100 uStim trials with the trained stimulation-response inverse model and one of the $R$ stimulation-response models. As a target state, we used one of the target neural population activity states planned to be used in the experiment with a monkey. On each trial, the simulation system randomly sampled one pre-uStim state from the $R$ experimental session datasets and passed it to the stimulation-response inverse model together with the target state. The stimulation-response inverse model then returned the stimulation suitabilities for each candidate electrode and chose the uStim pattern to perform using the same method described in Section~\ref{pattern_selection_method}. The sampled pre-uStim state and selected uStim pattern were then passed to the stimulation-response model to observe the expected post-uStim state. Finally, the pre-uStim state, the selected uStim pattern, and the predicted post-uStim state were accumulated to perform the PPO update of the stimulation-response inverse model as described in Section~\ref{ppo_method}. 

For each combination of hyperparameters, we performed $R \times T$ simulations (where each of the $R$ stimulation-response models was used to produce predicted post-uStim states for each of $T$ different targets) and computed the average absolute error between the predicted post-uStim state and target state over all simulations. We performed Bayesian Optimization with this error objective and decided the hyperparameter values based on the hyperparameter combination that achieved the smallest error. We run Bayesian Optimization using Optuna \citep{optuna_2019}. The ranges of hyperparameter values explored during Bayesian Optimization are summarized in the table below.

\begin{table}[htbp]
  \centering
\begin{tabular}{lc c c c cl}\toprule
            Method & Parameter & Description & Value \\\midrule
\midrule
\ uStim pattern selection &  $\varepsilon_\text{init}$ & Initial value of $\varepsilon_{c,k}$ & 0.3-0.7\\
\  & $\varepsilon_\text{floor}$ & Minimum value of $\varepsilon_{c,k}$ & 0.05-0.15\\
\  & $\gamma$ & Discount factor of $\varepsilon_{c,k}$ & 0.95-0.99\\
\ PPO & $\varepsilon_\text{clip}$ & Clip value of PPO update & 0.1-0.3\\
\  & $b$ & Baseline for advantage computation & 0.5-1.5\\
\  & - & PPO update frequency (trial) & 3-10\\
\  & - & Learning rate & 0.00001-0.001
\\\bottomrule
\end{tabular}
\end{table}


\clearpage

\begin{figure}[!tb]
\centering
\includegraphics[scale=0.9]{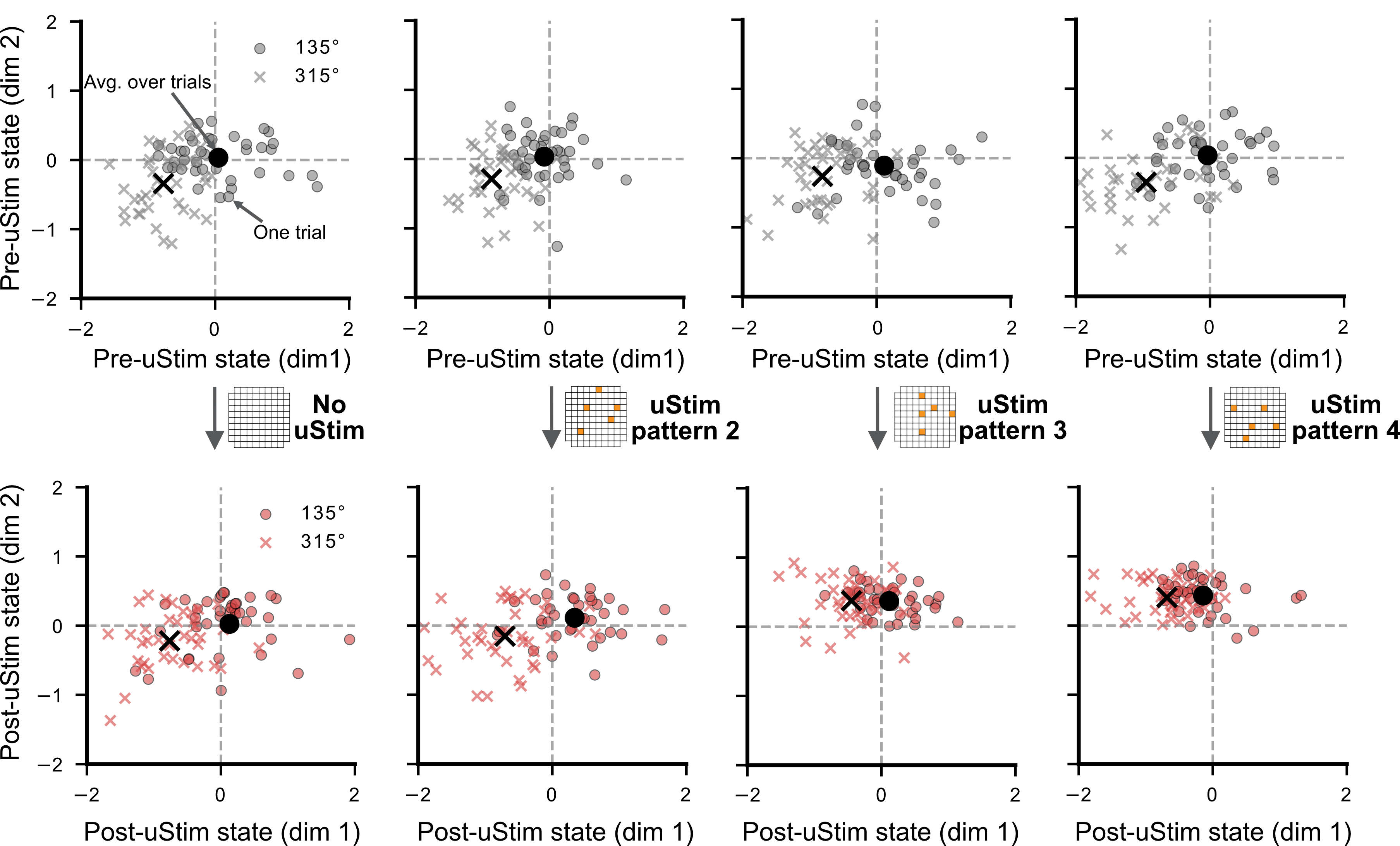}
\caption{\textbf{Pre-uStim states and subsequent post-uStim states induced by example uStim patterns.} Pre-uStim and post-uStim states with additional uStim patterns tested during the same experimental sessions with \cref{pre_vs_post_fig}A. Same conventions as in \cref{pre_vs_post_fig}A.}
\label{pre_vs_post_supfig}
\end{figure}

\end{document}